\documentclass[sigconf,nonacm]{acmart}

\AtBeginDocument{%
  }

\setcopyright{acmlicensed}
\copyrightyear{2025}
\acmYear{2025}
\acmDOI{XXXXXXX.XXXXXXX}

\acmConference[WWW ’25 Companion]{the ACM Web Conference 2025}{April 28-- May 02,
  2025}{Sydney, Australia}

\acmISBN{978-1-4503-XXXX-X/18/06}

\begin{document}

\title{ProxyLLM : LLM-Driven Framework for Customer Support Through Text-Style Transfer}

\author{Sehyeong Jo}
\affiliation{%
  \institution{University of Colorado Boulder}
  \city{Boulder}
  \state{CO}
  \country{USA}
}
\email{se.jo@colorado.edu}

\author{Jungwon Seo}
\affiliation{%
  \institution{University of Stavanger}
  \city{Stavanger}
  \country{Norway}}
\email{jungwon.seo@uis.no}

\renewcommand{\shortauthors}{Sehyeong Jo \& Jungwon Seo}


\renewcommand{\abstractname}{ABSTRACT}

\begin{abstract}
Chatbot-based customer support services have significantly advanced with the introduction of large language models (LLMs), enabling enhanced response quality and broader application across industries. However, while these advancements focus on reducing business costs and improving customer satisfaction, limited attention has been given to the experiences of customer service agents, who are critical to the service ecosystem. A major challenge faced by agents is the stress caused by unnecessary emotional exhaustion from harmful texts, which not only impairs their efficiency but also negatively affects customer satisfaction and business outcomes. In this work, we propose an LLM-powered system designed to enhance the working conditions of customer service agents by addressing emotionally intensive communications. Our proposed system leverages LLMs to transform the tone of customer messages, preserving actionable content while mitigating the emotional impact on human agents. Furthermore, the application is implemented as a Chrome extension, making it highly adaptable and easy to integrate into existing systems. Our method aims to enhance the overall service experience for businesses, customers, and agents.
\end{abstract}

\ccsdesc[500]{Information systems~Web applications}
\ccsdesc[500]{Information systems~Web services}
\ccsdesc[500]{Human-centered computing~Natural language interfaces}

\keywords{Large Language Models, Human-centered Interfaces, Sentiment Analysis}

\maketitle

\begin{figure*}
    \centering
    \includegraphics[width=\textwidth]{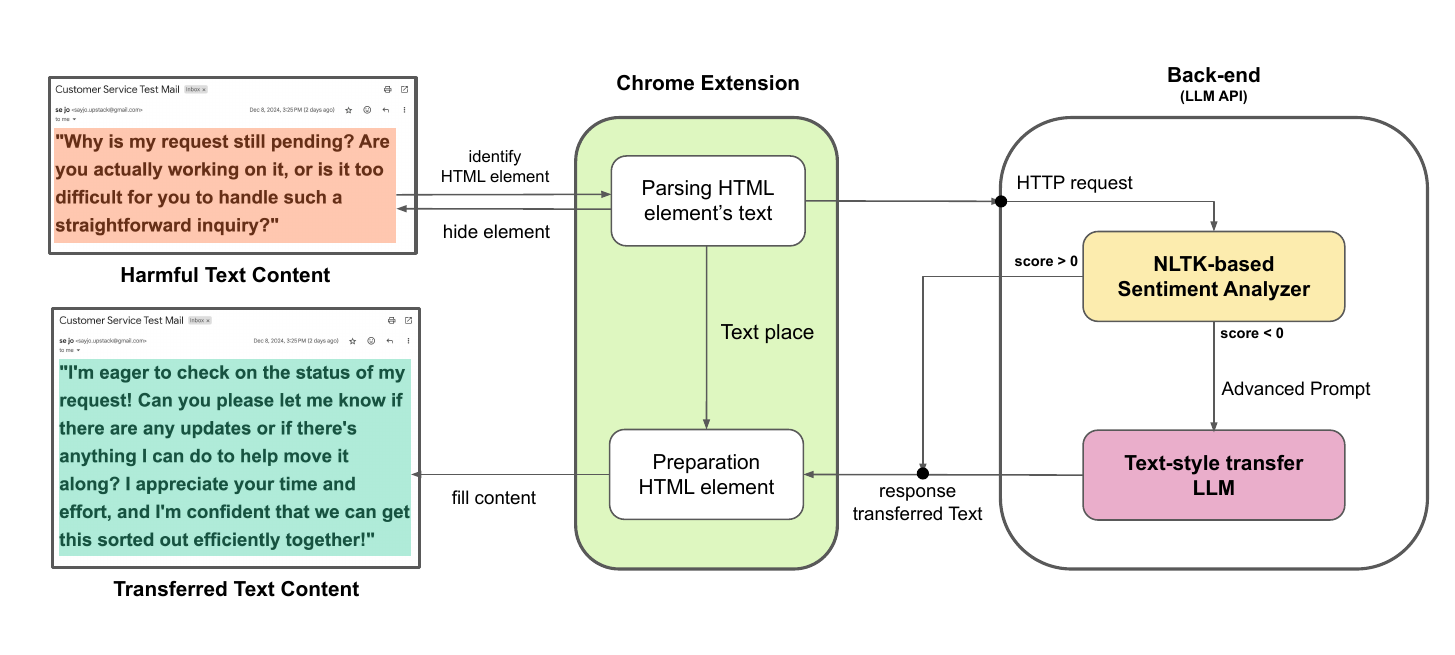}
    \caption{Overview of the ProxyLLM Application System.}
    \label{fig:overview}
\end{figure*}

\section{INTRODUCTION}
Chatbot-based customer support services have become increasingly prevalent across industries~\cite{tian2023multi,cui2017superagent,xu2017new}. Companies adopt these automated services to reduce operational costs and improve efficiency. Driven by advancements in large language models (LLMs), these systems have surpassed traditional chatbots in performance, offering more accurate and context-aware responses~\cite{dam2024complete}. Consequently, LLM-based chatbots are being widely adopted for their ability to enhance customer satisfaction and address diverse scenarios across various sectors.

Despite advances in automation, human agents continue to play a critical role in managing complex interactions~\cite{banerjee2023system,zhou2023webarena}. These interactions, however, often impose emotional strain on agents, as customers may include emotionally charged or negative comments alongside their inquiries or requests. Human agents are bound by company guidelines to respond professionally and within strict rules, regardless of the emotional nature of customer messages. This often means addressing both the actionable content and the emotional undertone of the message, which can be draining over time~\cite{bellet2024does}. Addressing this challenge requires rethinking the interaction pipeline.

Therefore,  we propose ProxyLLM, a system that functions similarly to a proxy server, positioned between human agents and existing systems to prevent human agents from being directly exposed to raw harmful messages. Specifically, we propose using LLMs to adjust the tone of incoming messages using text-style-transfer~\ref{toshevska2021review} while preserving their core content. This adjustment aims to shield agents from unnecessary emotional distress, enabling them to work more efficiently and with greater composure. Furthermore, to ensure ease of use and seamless integration, ProxyLLM is implemented as a Chrome extension\footnote{\url{https://developer.chrome.com/docs/extensions}}. This implementation enables effortless integration into diverse web-based customer service systems without necessitating modifications to existing system pipelines or the deployment of additional components, such as databases. Notably, the integrity of the original messages remains preserved. 

Our contributions are summarized as follows:

\begin{itemize}
    \item We introduce a novel application leveraging LLMs to enhance the mental health support for human agents in high-stress environments.
    \item By employing the style-transfer capabilities of LLMs, our approach preserves the critical content of communications while effectively filtering harmful or distressing material before delivery to the agents.
    \item We implement the system as a lightweight Chrome extension, ensuring seamless integration with existing workflows without introducing conflicts.
\end{itemize}

Our source code and related materials can be found at: ~\url{https://github.com/sayjoupstack/Proxy-LLM}.

\section{ProxyLLM}
This section provides an overview of the architecture and functionality of the proposed method, ProxyLLM.

\subsection{System Overview}
The architecture of the proposed system consists of two primary components: (1) a Chrome extension and (2) a back-end machine learning model server, as depicted in Figure~\ref{fig:overview}.

\noindent\textbf{Chrome Extension.} A Chrome extension is a third-party application designed to enhance browser functionality, supporting tasks such as ad-blocking and grammar correction~\cite{hsu2024chrome}. It is packaged as a compressed archive containing HTML, JavaScript, and CSS files, with operations governed by user-granted permissions. The Chrome extension for ProxyLLM is developed to automatically detect HTML elements containing potentially harmful text and obscure these elements to shield human agents from exposure. It also enables seamless transmission of the identified text to a back-end API for text-style transformation. Additionally, the extension serves as a graphical user interface (GUI), allowing human agents to fine-tune custom prompts for style-transfer tasks, ensuring tailored and contextually appropriate responses.

\noindent\textbf{Back-End Processing.} The back-end component processes incoming requests using a Natural Language Toolkit (NLTK)~\cite{bird2006nltk} sentiment analysis module~\cite{shelar2018sentiment}, which computes sentiment scores ranging from -1 to 1. This lightweight sentiment analyzer minimizes computational overhead by bypassing the LLM when sentiment scores fall outside predefined thresholds. For requests requiring style transfer, the custom prompts (from Chrome extension) undergo processing through a text-style transfer module, leveraging straightforward prompting techniques~\cite{oppenlaender2024prompting,reif2021recipe}. The transformed text is then transmitted back to the Chrome extension, where it is seamlessly integrated into the appropriate HTML elements within the agent’s interface. 
\subsection{Implementation Detail}
The application’s back-end architecture leverages Flask, providing a lightweight yet robust API system that ensures efficient communication between the Chrome extension and server-side operations. Sentiment analysis is performed using the NLTK, utilizing streamlined algorithms to deliver rapid and accurate evaluations. This approach minimizes computational overhead, avoids latency issues associated with more complex ML models, and ensures compatibility with resource-constrained environments.

Text-style transfer tasks are powered by the Llama 3.1 8B model~\cite{dubey2024llama}, chosen for its optimal performance under limited GPU resources, facilitating scalability and adaptability across various deployment scenarios. The Ollama framework\footnote{\url{https://ollama.com/}} further enhances flexibility, enabling service providers to dynamically adjust the underlying model to meet specific system requirements and performance needs. Demonstrations are conducted on an RTX 2060 GPU, illustrating the system’s capability to function effectively under constrained computational resources. This architecture supports tailored solutions, ensuring ProxyLLM can adapt to diverse operational constraints.

\subsection{Prompt Customization}
\subsubsection{Sentiment Setting}

\begin{figure}
    \centering
    \includegraphics[width=\linewidth]{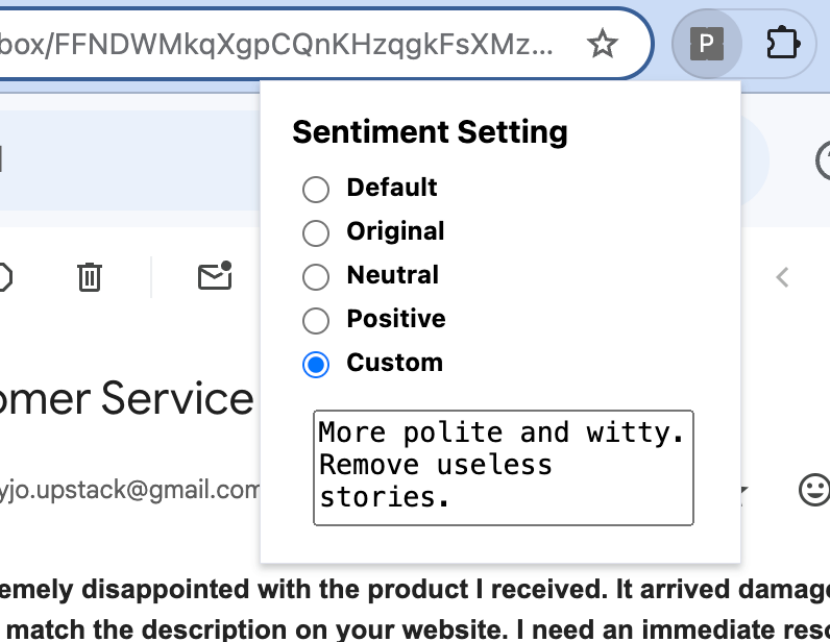}
    \caption{Personalized Sentiment Setting}    
    \label{fig:sent-set}
\end{figure}

This system includes predefined sentiment presets—original, neutral, positive, and custom—to provide flexibility in tone adaptation~(Figure~\ref{fig:sent-set}). For the custom level, agents can design tailored prompts to define the tone and structure of displayed content, enabling precise alignment with specific communication needs.

\subsubsection{Prompting}

Text-style transfer can be achieved through two approaches: leveraging pre-trained ML models~\cite{madaan2020politeness} or utilizing LLMs~\cite{luo2023prompt}. ProxyLLM adopts the latter approach, employing LLMs to generate results through prompting, thereby reducing the cost of additional training while enabling customization.

The basic method for generating transformed text involves appending a directive to rephrase the original content with enhanced politeness. The basic prompt structure is :

\begin{center}
    \vspace{6pt}
    \fbox{
        \begin{minipage}{0.9\linewidth}
            \emph{This is original text. Change this text style. \textbf{[Original Text]}. Change this text content to be more polite.}
        \end{minipage}
    }
    \vspace{6pt}
\end{center}

In this case, the LLM determines the level of sentiment based on its inherent standards, which are reflected in the transformed text. However, relying solely on a basic prompt leaves the criteria for sentiment entirely dependent on the LLM's choices~\cite{sahoo2024systematic}, without reflecting personal customization.

\begin{figure}
    \centering
    \includegraphics[width=\linewidth]{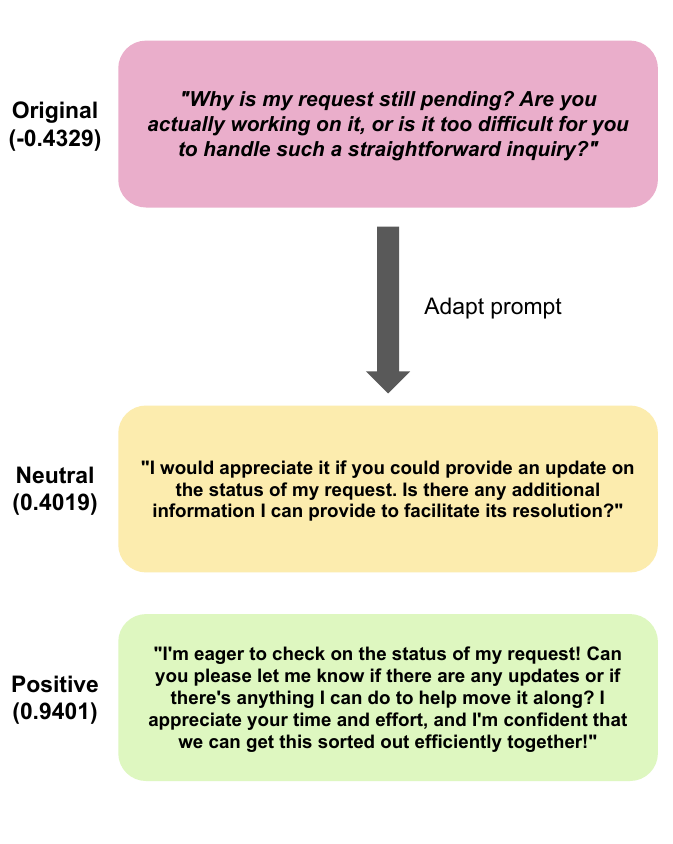}
    \caption{The results for neutral and positive sentiment, along with examples of their respective sentiment scores, after applying a custom advanced prompt to the original text.}
    \label{fig:sent-prompt}
\end{figure}

To provide customized results to human agents, ProxyLLM integrates personalized configuration parameters into the prompt~(Figure~\ref{fig:sent-prompt}). The advanced prompt structure is:

\begin{center}
    \vspace{6pt}
    \fbox{
        \begin{minipage}{0.9\linewidth}
            \emph{This is original text. Change this text style. \textbf{[Original Text]}. Change this text content to be more \textbf{[Custom Parameter]}.}
        \end{minipage}
    }
    \vspace{6pt}
\end{center}

The original text is modified in alignment with the personalized configuration parameters. If the parameter corresponds to predefined presets, such as "neutral" or "positive," a pre-designed prompt tailored to each preset is applied. For custom parameters, a tailored prompt is employed to achieve the desired transformation. An example of a prompt corresponding to a custom parameter is as follows:

\setlength{\itemsep}{3pt}
\begin{description}
  \item[$\bullet$ Neutral :] \textit{rewriting in a neutral tone to remove any emotional, biased, or subjective elements while preserving the original meaning.}
  \item[$\bullet$ Positive :] \textit{rewriting in a positive tone, enhancing the optimism and uplifting language while preserving the original meaning and intent.}
\end{description}

\subsubsection{Evaluation}
To evaluate the effectiveness of ProxyLLM's text style transfer, we conducted an assessment by treating commercial LLMs (GPT-4o-mini~\footnote{\url{https://chatgpt.com/}}, Claude Sonnet~\footnote{\url{https://claude.ai/}} and Gemini~\footnote{\url{https://gemini.google.com/}} ) as if they represented real individuals. We prepared a dataset~\footnote{The dataset is available in the GitHub repository.} comprising 10 sets of original and transferred texts. The original texts are manually crafted inquiry samples with a negative tone, while the transferred texts are generated using ProxyLLM's Llama 3 with the custom setting adjusted to \textit{positive}. The structure of the prompt for requesting the score is as follows:

\begin{center}
    \vspace{6pt}
    \fbox{
        \begin{minipage}{0.9\linewidth}
            \emph{Please mark the sentimental score I present from -1.0 (very negative) to 1.0 (very positive). \textbf{[Original Text]}.
            }
        \end{minipage}
    }
    \vspace{6pt}
\end{center}

\begin{table}[h!]
    \centering
    \caption{The average sentiment score variations original and transferred applying LLMs to sample texts.}
    \begin{tabular}{c|c|c}
        \hline
        \textbf{LLMs} & \textbf{Original} & \textbf{Transferred} \\ 
        \hline
        GPT-4o-mini & -0.48 & 0.28\\ 
        Claude Sonnet & -0.66 & 0.24\\ 
        Gemini & -0.6 & 0.18\\ 
        \hline
        NLTK & -0.48 & 0.83\\ 
        \hline
    \end{tabular}
    \label{table:shift-table}
\end{table}

The evaluation results presented in Table~\ref{table:shift-table} reveal several key insights. Firstly, texts transformed by Llama3 consistently received positive evaluations across all tested LLMs. These findings confirm that using a locally hostable LLM combined with a straightforward prompt enables the achievement of consistent tone conversion as desired. Furthermore, the NLTK-based sentiment analysis, introduced to reduce computational overhead, proves to be reliable. This reliability is substantiated by the minimal discrepancy between the original NLTK scores and the sentiment scores derived from the original texts of other LLMs, underscoring its capability to effectively detect negative tonalities.

\section{APPLICATION WORKFLOW}

\begin{figure}[htbp]
    \centering
    \includegraphics[width=\linewidth]{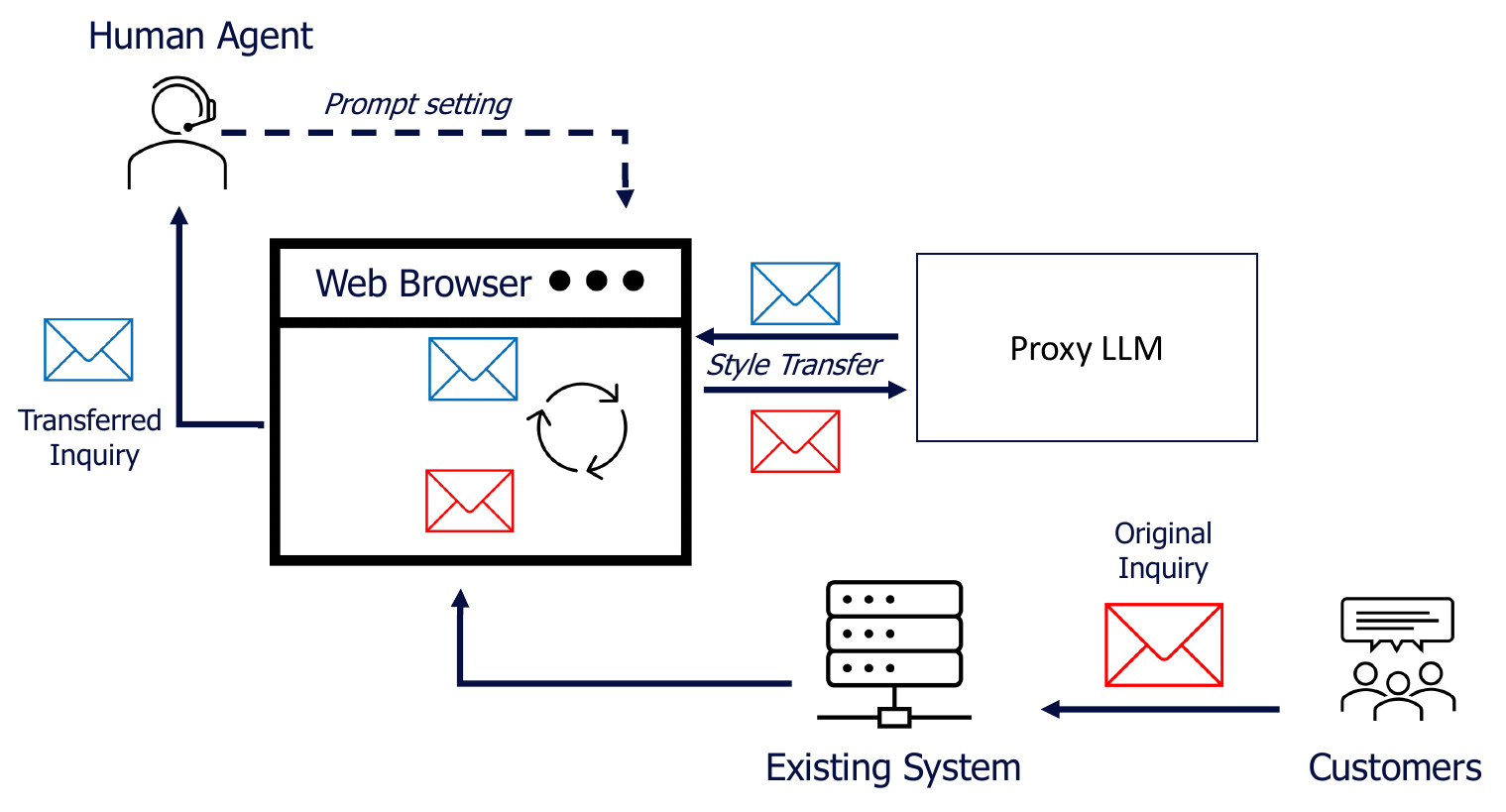}
    \caption{ProxyLLM Workflow Diagram}    
    \label{fig:workflow}
\end{figure}
As shown in Figure~\ref{fig:workflow}, ProxyLLM operates passively without requiring a dedicated program to be launched. When a customer inquiry is displayed on a web page, the text content within the relevant section is temporarily masked. Once the processing is complete, the transformed text received from the server replaces the original content in that section. For custom settings, users can click the corresponding Chrome extension icon to select a desired preset. If a custom prompt is needed, it must be applied by selecting the "Custom" option. However, if the "Custom" option is chosen without providing a prompt, the system defaults to using the standard preset, delivering results based on the default configuration. The current demo version of ProxyLLM is configured specifically for the Gmail environment. Future feature developments aim to enhance adaptability by enabling modifications to the designated HTML regions of text content. This enhancement would allow ProxyLLM to be applied to other email services, such as Outlook, as well as various other web services.

\section{CONCLUSION}

In this study, we introduced ProxyLLM for customer support by leveraging the capabilities of LLM to address the complementary objectives of improving service quality and protecting the mental well-being of human agents. By effectively transforming harmful or emotionally charged customer messages into neutral or positive tones, the system ensures a more constructive interaction environment for customer service agents. Furthermore, its seamless integration as a Chrome extension demonstrates the practical and adaptable design of ProxyLLM, enabling it to enhance existing customer support workflows without incurring additional infrastructure costs.

Future research on ProxyLLM could explore its potential beyond the current application, given its modular design that facilitates adaptation to other web services and platforms. Future work could explore enhancements in personalization, model efficiency, and broader multilingual capabilities to address diverse customer needs globally. By combining advanced text-style transfer techniques with an agent-centric approach, ProxyLLM provides a framework for developing AI-driven solutions that enhance the human aspects of service delivery.

\bibliographystyle{ACM-Reference-Format}
\bibliography{sample-base}

\end{document}